\documentclass[fleqn,usenatbib]{mnras}

\DeclareRobustCommand{\VAN}[3]{#2}
\let\VANthebibliography\thebibliography
\def\thebibliography{\DeclareRobustCommand{\VAN}[3]{##3}\VANthebibliography}

\usepackage{graphicx}	% Including figure files

\title[Effects of Data Shortage on Solar Activity Forecast]{Effects of Observational Data Shortage on Accuracy\\of Global Solar Activity Forecast}

\author[I. N. Kitiashvili]{
Irina N. Kitiashvili \thanks{E-mail: Irina.N.Kitiashvili@nasa.gov}
\\
NASA Ames Research Center, Moffett Field Blv, MS 258-5, Mountain View, 94035, USA}

\date{Accepted XXX. Received YYY; in original form ZZZ}

\pubyear{2020}

\begin{document}
\label{firstpage}
%\pagerange{\pageref{firstpage}--\pageref{lastpage}}
\maketitle

% Abstract of the paper
\begin{abstract}
Building a reliable forecast of solar activity is a long-standing problem that requires an accurate description of past and current global dynamics. Relatively recently, synoptic observations of magnetic fields and subsurface flows have become available. In this paper, we present an investigation of the effects of short observational data series on the accuracy of solar cycle prediction. This analysis is performed using the annual sunspot number time-series applied to the Parker-Kleeorin-Ruzmaikin dynamo model and employing the Ensemble Kalman Filter (EnKF) data assimilation method. The testing of cycle prediction accuracy is performed for the last six cycles (from Solar Cycle 19 to 24) by sequentially shortening the observational data series that are used for prediction of a target cycle and evaluating the resulting prediction accuracy according to specified criteria. According to the analysis, reliable activity predictions can be made using relatively short time-series of the sunspot number. It is demonstrated that even two cycles of available observations allow us to obtain reasonable forecasts.
\end{abstract}

% Select between one and six entries from the list of approved keywords.
% Don't make up new ones.

%%%%%%%%%%%%%%%%%%%%%%%%%%%%%%%%%%%%%%%%%%%%%%%%%%

%%%%%%%%%%%%%%%%% BODY OF PAPER %%%%%%%%%%%%%%%%%%

\section{Introduction}
Forecasting of the global solar activity characterized by sunspot number (SN) variations on the 11-year time scale is a challenging problem, principally due to our limited knowledge about physical processes hidden under the solar surface. A number of approaches developed to predict the basic properties of the current solar cycle have produced inconsistent results \citep[see review by][]{Pesnell2012} due to uncertainties in models and observations. 

Thus, the goal of the paper is to test the performance of the data assimilation procedure applied to the PKR model in a case of limited knowledge about past solar activity. The long length of the sunspot observations makes it possible to test the capabilities of this procedure to predict stronger, weaker, or moderate cycles for a different range of available observations, and develop criteria to evaluate the resulting predictions. This step is critical for the transition to the more sophisticated models, for which the ability to perform intensive testing is limited.

We perform test predictions from Cycle 19 (SC19) to Cycle 24 (SC24). In the next section, we describe the annual sunspot number data series used in the analysis. Sections~\ref{sec:MFmodel} and~\ref{sec:method} include a brief description of the Parker-Kleeorin-Ruzmaikin (PKR) mean-field dynamo model and methodology. Section~\ref{sec:tests} presents reconstructions for six target cycles using the sunspot number observations of different lengths preceding these cycles. We discuss the accuracy of the predictions in Section~\ref{sec:Discussion}, and summarize our conclusions in Section~\ref{sec:Conclusions}.

\section{Sunspot number time-series}
\begin{figure}
	\includegraphics[width=\columnwidth]{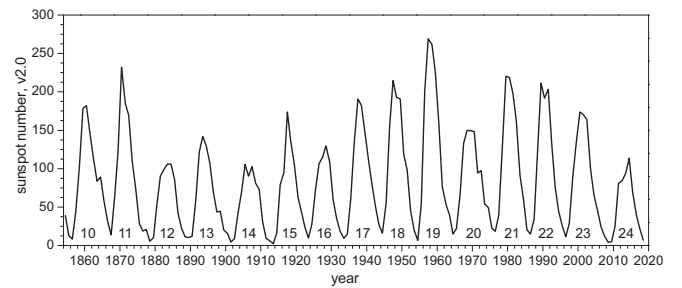}
	\caption{The annual sunspot number (v2.0) time-series from the WDC-SILSO repository of the Royal Observatory (Brussels, Belgium).}
	\label{fig:sunspot}
\end{figure}

We use the annual sunspot number time-series data (version 2.0, Fig.~\ref{fig:sunspot}) from the WDC-SILSO{\footnote{http://www.sidc.be/silso/datafiles}} repository of the Royal Observatory (Brussels, Belgium). According to our previous studies, the different calibrations (or versions) of the sunspot number series do not affect the analysis results after applying the corresponding observational data rescalings \citep{Kitiashvili2016}.

\section{Mean-field dynamo model}\label{sec:MFmodel}
For the data assimilation procedure, we use the mean-field dynamo model described by \cite{Kitiashvili2008a,Kitiashvili2009}, in which the Parker's migratory  $\alpha\Omega$-dynamo model \citep{Parker1955} is combined with the equation of magnetic helicity balance \citep{Kleeorin1982,Kleeorin1995}.
It has been shown that taking into account the evolution of the magnetic helicity provides an explanation of Waldmeier's relations \citep{Kitiashvili2008a,Pipin2011}. In a simplified form, the dynamo model is described by the following equations: 
\begin{eqnarray}
\frac{\partial A}{\partial t}&=&\alpha B+\eta \nabla ^{2}A \nonumber \\
\frac{\partial B}{\partial t}&=&G\frac{\partial A}{\partial x}+\eta
\nabla ^{2}B,\\
\frac{\partial \alpha_{m}}{\partial t}&=&-\frac{\alpha_{m}}{T}+\frac{Q}{4\pi \rho}\left[\left<{\bf B}\right>  \cdot
(\nabla \times \left<{\bf B}\right>)-\frac{\alpha}{\eta}\left<{\bf B}\right>^{2}\right], \nonumber
\end{eqnarray}
where $A$ and $B$ are the vector-potential and the toroidal field component of magnetic field $\rm{\bf B}$, $\alpha=\alpha_{h}+\alpha_{m}$ is total helicity, which includes the hydrodynamic helicity ($\alpha_{h}$), and magnetic helicity ($\alpha_{m}$) parts, $\eta$ is the total magnetic diffusivity, which includes the turbulent and molecular magnetic diffusivities, $\eta=\eta_{t}+\eta_{m}$ (usually $\eta_{m} \ll \eta_{t}$), $G$ describes the rotational shear, $\rho$ is density, $Q$ and $T$ are coefficients which describe characteristic scales and times \citep{Kleeorin1982}, $x$ is the radial coordinate, and $t$ is time.

To perform the data assimilation analysis, we employ the low-mode approximation \citep{Weiss1984}, which reduces the mean-field Parker-Kleeorin-Ruzmaikin (PKR) dynamo model (Eq. 1) to the following non-linear dynamical system \citep{Kitiashvili2009,Kitiashvili2020a}

\begin{eqnarray}
\frac{{\rm d} A}{{\rm d} t}&=& D B- A,  \qquad
\frac{{\rm d} B}{{\rm d} t}={\rm i} A - B, \nonumber \\
\frac{{\rm d} \alpha_{m}}{{\rm d} t}&=&-\nu \alpha_{m} - D \left[
B^{2}-\lambda A^2 \right],
\end{eqnarray}
where $A$, $B$, $\alpha_{m}$ and $t$ are non-dimensional variables, $D=D_{0}\alpha$ is the non-dimensional dynamo number, $D_{0}=\alpha_{0}Gr^{3}/\eta^{2}$, $\alpha_{0}=2Qk\upsilon_{A}^{2}/G$, where $\upsilon_{A}$ is the Alfv\'en
speed, $\lambda=(k^2\eta/G)^2={\rm R_m^{-2}}$, $k$ is a characteristic wavelength, $\nu\sim 2 k_0 \eta_m^2 /(k^2\eta_t)$, and $k_0$ is the largest scale in the inertial range of the turbulence spectrum \citep{Kleeorin1982}.

Despite the simple formulation, the model describes the basic features of solar cycles: generation and transformation of the poloidal and toroidal magnetic field components as well as the evolution of the sunspot number, $W$, which is modeled following the Bracewell's relation $W = {\rm const} B^{3/2}$.  It is important to note, that including effects of the magnetic helicity balance has demonstrated improvement of the 2D mean-field dynamo model to reproduce the sunspot cycles properties \citep{Pipin2011}.
Previous studies demonstrated the ability of the mean-field PKR model to reproduce the qualitative properties of the  sunspot cycles \citep{Kitiashvili2009}, thus making the model suitable for application with the Ensemble Kalman Filter. 

\section{Assimilation of the sunspot number time series into the PKR dynamo model using Ensemble Kalman Filter}\label{sec:method}
The previous successful application of the data assimilation approach \citep{Kitiashvili2008a}, and, in particular, the Ensemble Kalman Filter method \citep[EnKF,][]{Evensen1994}, can be explained by the following factors: 1) the derived mean-field dynamo model is able to reproduce the general properties of sunspot cycles; 2) the model and assimilated data are compatible in terms of the complexity and robustness; 3) the availability of long data-series of the sunspot number is critical for the model calibration and estimation of observational uncertainties. According to the dynamo theory  \citep{Babcock1961,Leighton1969}, active regions represent toroidal magnetic fields generated in the solar interior as a result of differential rotation. The qualitative relationship between the toroidal magnetic field component and the sunspot number, $W = {\rm const} B^{3/2}$ (where $W$ is the sunspot number, and $B$ is the toroidal field component) suggested by \cite{Bracewell1988}, provides a link between the sunspot number time series and the general global field variations. In this paper, we use this relationship to assimilate the available sunspot number time-series into the reduced dynamo model (Eq. 2). For performing the EnKF analysis, the periodic model solution is normalized to the strength of the last observed solar cycle in our test series. In the case of a phase discrepancy between the model and data by two or more years, the model solution is shifted to get a better phase agreement of the solar cycle minimum times. Synthetic observations of the poloidal field and magnetic helicity are generated from the model solution with added random noise of $10\%$. 

\section{Tests of solar cycle prediction using a limited amount of sunspot data}\label{sec:tests}
To investigate the influence of short observational data series, we perform a series of test predictions for six solar cycles from 19 to 24. Including SC19 and SC21 allows us to test the ability to predict transition to stronger activity cycles. We perform a sequence of test predictions which use 9, 8, ..., 2 preceding cycles of the sunspot observations; we use 300 ensemble members in the EnKF procedure in all cases. For example, for the test prediction of Solar Cycle 19 (SC19), we initially use a nine-cycle time-series of sunspot number, SC10 to SC18. Then, we progressively remove cycles of observations from the analysis, starting with SC10, and sequentially evaluate the cycle prediction accuracy.

The performed prediction tests led us to the following criteria to examine the quality of the obtained solar activity predictions, in descending order of importance:
\begin{itemize}
	\item[1.] The phase discrepancy between the exact model solution and observations should not be greater than 2 years.
	\item[2.] If at the end of the reconstruction phase the exact solution of the toroidal magnetic field component is near zero, it can be considered as evidence of high prediction accuracy.
	\item[3.] The signs of the last available observation for the toroidal field and the corresponding model solution should be the same; however, if the last observational data point is too far from the solar minimum it can significantly reduce the forecast accuracy.
	\item[4.] The exact model solution for the prediction phase must be consistent with the model solution for the reconstruction phase (the solution should not show extreme flattening or exhibit jumps or bumps, but the solution may be shifted according to the new initial condition).
	\item[5.] The corrected solution (first guess estimate) at the initial time during the prediction phase should not be greater than the best-estimate variations of the toroidal field. This criterion provides only a suggestion about the possibility of an inaccurate prediction, therefore it is applicable if the previous criteria are satisfied.
\end{itemize} 
The criteria to evaluate the accuracy of a forecast are defined from the prediction performance that was obtained for the different lengths of the observational time-series. The forecasts are rated as accurate if the maximum sunspot number is predicted with the amplitude error not exceeding 15\%, and the time error not exceeding 1 year. 

The first three criteria are most important for the evaluation of a prediction. The first criterion was previously identified by \cite{Kitiashvili2008a}. The second one comes from the previously found importance of setting up the initial conditions to perform a prediction of the upcoming activity during periods of a high ratio between the toroidal and poloidal magnetic field components \citep{Kitiashvili2016}. The third criterion intends to capture potentially strong discrepancies between the model solutions and the last available observation of the toroidal field. The fourth criterion captures the abnormal behavior of the non-linear solution in the rising phase of the predicted cycle, such as the appearance of unexpected fluctuations of the toroidal field. The last, fifth, criterion can be used only if several prediction results more or less similarly satisfy the previous four criteria and provide a broad range of predicted cycle properties. This primarily happens if the observational time series is only 2--3 cycles long.

According to the study, the most efficient approach in obtaining an accurate forecast includes the following steps:
\begin{itemize}
	\item Perform a test forecast of a past cycle to calibrate the phases of the periodic model solution and preceding solar activity to obtain agreement between the predicted and actual cycle properties. 
	\item Repeat the previous analysis to perform prediction of the next cycle by using information about the model solution phase. If the discrepancy between the last observed cycle and the model solution is more than two years, shift the model solution by one year.  
\end{itemize} 
This sequential approach has been previously used to predict Solar Cycle 24 \citep{Kitiashvili2008a}. 

Following the criteria described above, the test prediction results and the corresponding solutions for the toroidal magnetic field component for Solar Cycles 19 -- 24, for the various lengths of sunspot time series, are shown in the Appendix. Below we describe the prediction accuracy for each of the six test cycles.

{\bf Solar Cycle 19} is the strongest in the test series. The predicted and observed SC19 sunspot numbers are in agreement when the EnKF analysis uses sunspot time-series 9, 7, and 4 cycles long (Fig.~\ref{fig:SC19}). Evaluation of the toroidal field variations (Fig.~\ref{fig:SC19er}) in most cases shows the same sign for both the last observed toroidal field (estimated from the sunspot number according to the Hale law) and the model solution (black lines). According to the above criteria, the obtained predictions are likely to be accurate. However, only these three of the eight test predictions show good agreement with the actual observations.

{\bf Solar Cycle 20} is significantly weaker than SC19. In the forecast, the previous prediction results for SC19 have been taken into account. For instance, for the prediction of SC20, in which the observational data set was 5-cycles long, the phase of the periodic model solution was taken from the prediction of SC19, which was obtained with 4-cycles of observations and was accurate. In most cases, the predicted accuracy of the SC20 forecast, based on the properties of the toroidal field variations (Fig.~\ref{fig:SC20er}), corresponds to the actual observations (Fig.~\ref{fig:SC20Pred9-2cycles}). Only three predictions were incorrectly rated by our {\it a priori} criteria as accurate. Nevertheless, the rising phase of solar activity up to the solar maximum was reproduced correctly in these cases.  

{\bf Solar Cycle 21} is 50~\% stronger than the previous cycle, SC20. According to the forecast accuracy criteria, seven of the eight test predictions were expected to be accurate. The test using two cycles of observation was rated as not sufficiently accurate. However, even in this case, the prediction results follow the actual sunspot data. It is important to note that, in some cases that are rated as accurate according to our defined criteria, there may nevertheless be noticeable discrepancies in the toroidal field and sunspot observations during the rising phase of activity (Fig.~\ref{fig:SC21Pred9-2cycles},~\ref{fig:SC21er}).
In this and the following analyses, we used our test results for predictions of the previous cycles.

{\bf Solar Cycle 22} predictions showed performance similar to SC21. All forecasts were expected to provide accurate estimations of the strength and time of the solar maximum (Fig.~\ref{fig:SC22Pred9-2cycles},~\ref{fig:SC22er}). However, in the test case with 4 preceding cycles, solar activity during SC22 was underestimated by 23\%.

{\bf Solar Cycle 23} followed the trend of long-term decreasing global activity and was weaker than the previous cycle by 20\%. Evaluation of the toroidal field evolution allowed us to correctly identify all accurate and inaccurate predictions  (Figs.~\ref{fig:SC23Pred9-2cycles},~\ref{fig:SC23er}). 

{\bf Solar Cycle 24} is the weakest in the considered time-series with a maximum sunspot number of about 60\% smaller than the previous activity cycle. 
The forecast quality control correctly identified five accurate predictions and one inaccurate. Two cases of overestimated cycle strengths (by 31\% and 22\%) were identified correctly.

\section{Discussion}\label{sec:Discussion}
Future development of tools for solar activity prediction requires understanding the limitations caused by short time-series of observational data. The first step in this direction is performed in this study, in which the sunspot number series of various lengths are assimilated into a simplified dynamo model using the EnKF method. The test prediction results for the last six cycles using different lengths of the preceding data-series are presented in Section~\ref{sec:tests} and Appendix. In this study, the sunspot number is converted into the toroidal magnetic field component following the three-halves law suggested by \cite{Bracewell1988}, with proper scaling and sign definition according to Hale's law. However, there is some freedom to choose which observations are considered as the first and last in the observational data set. Thus it is possible to obtain several forecasts using essentially the same amount of observational data, which raises the question of evaluating these predictions to choose the best one.

We rank our test predictions using the criteria listed in the previous section. These tests are evaluated for forecast accuracy, and thus for our ability to evaluate if the predicted strength of a target cycle is correct. If according to our criteria we expect that a prediction result is accurate, and if the predicted sunspot maximum indeed deviates from the observed value by less than 15\%, we consider this test result as a true positive (TP). If we expect an accurate prediction, but the predicted maximum of the sunspot number deviates from the actual data by more than 15\%, then the test case is identified as false positive (FP). Similarly, if the obtained solution does not follow our first three criteria, we consider this forecast as inaccurate. However, if the comparison of the predicted cycle strength with the actual sunspot maximum is less than 15~\%, the test case is ranked as false negative (FN). Similarly, we define the test as true negative (TN) for larger discrepancies between the prediction and observations. A statistical representation of the performance is shown in Table~\ref{tab:table}. Also, it is useful to include combinations of these scores such as {\it Accuracy} ((TP+TN)/(TP+FP+FN+TN)), {\it Precision} (TP/(TP+FP)), {\it Recall} (TP/(TP+FN)), and {\it F1-score} (2 $\times$ Precision $\times$ Recall /(Precision+Recall)). 

\begin{table}
	\centering
	\caption{Performance metrics of solar activity forecasts.}
	\label{tab:table}
	\begin{tabular}{lcccccc} % four columns, alignment for each
		\hline
           & SC19 & SC20 & SC21 & SC22 & SC23 & SC24\\
        \hline
           TP& 3 & 5 & 7 & 7 & 7 & 5  \\
           TN& 1 & 0 & 0 & 0 & 1 & 1  \\
           FP& 4 & 3 & 0 & 1 & 0 & 2  \\
           FN& 0 & 0 & 1 & 0 & 0 & 0  \\
%         \hline
Accuracy   & 0.500 & 0.625 & 0.875 & 0.875 & 1.000 & 0.625 \\
Precision  & 0.429 & 0.625 & 1.000 & 0.875 & 1.000 & 0.714 \\
Recall     & 1.000 & 1.000 & 0.875 & 1.000 & 1.000 & 1.000 \\	
F1-score   & 0.600 & 0.769 & 0.933 & 0.933 & 1.000 & 0.833 \\
       \hline
	\end{tabular}
\end{table}

As shown in Table 1, the {\it accuracy} of the prediction is the lowest for SC19 when no additional knowledge about the correspondence between the modeled and observed activity phases was used. Therefore, increasing accuracy for the other target cycles is expected because the previous tests with the assimilation using about the same amount of observational data allowed us to better calibrate our model. The {\it precision} characteristic, which means the ability to identify accurate predictions, also increases as we accumulate more information during the test predictions for the previous cycles. The {\it recall} characteristic describes the ability to pinpoint all correct predictions, shows good performance for all target cycles. 
Concerning the test results with assimilation of observational data only up to four preceding cycles, it is important to note that the incorrect ratings of forecast accuracy were only in three cases: 
the false-positive prediction of SC19 using two preceding solar cycles, the false-negative forecast for SC21, also using two cycles of observations, and the false-positive prediction of SC22 with 4 preceding cycles. The overall performance of our test predictions suggests that short time-series of observations can be successfully applied to make a reliable forecast.

\section{Conclusions}\label{sec:Conclusions}
The data assimilation approach of combining a global dynamo model with the annual sunspot number through the Ensemble Kalman Filter method has demonstrated great potential to make reliable forecasts for solar activity cycles. Further development of new physics-based capabilities to predict solar activity on global scales requires assimilation of synoptic observations, such as magnetic fields, into more detailed dynamo models. This raises a question about the dependence of the forecast accuracy on the length of the available observational data series. 

In this paper, we used the annual sunspot number to test the possibility of performing reliable forecasts for short time-series of observational data and developed a methodology to evaluate the accuracy of the obtained prediction.
A series of tests to predict the solar activity for six target cycles for different lengths of observed time-series showed the ability to build reliable long-term activity forecasts. In this work, the criteria to evaluate the forecast accuracy were developed. 
While the performed analysis demonstrates a good capability to produce reliable forecasts for stronger and weaker cycles, it also showed the importance of long time-series of observations. Longer time-series allow us to perform test predictions for the optimal choice of the model  parameters and to estimate the model uncertainty. We have expanded this study to the analysis of synoptic magnetograms, which are available for the last four cycles of solar activity \citep{Kitiashvili2020a}. Later steps will include development of data assimilation analysis for 2D observational data and a more detailed dynamo model.

\section*{Acknowledgements}
The work is supported by NSF grant AGS-1622341.

%%%%%%%%%%%%%%%%%%%%%%%%%%%%%%%%%%%%%%%%%%%%%%%%%%
%%%%%%%%%%%%%%%%%%%% REFERENCES %%%%%%%%%%%%%%%%%%

% The best way to enter references is to use BibTeX:

\bibliographystyle{mnras}
%\bibliography{assim_solar_cycle} % if your bibtex file is called example.bib

%%%%%%%%%%%%%%%%%%%%%%%%%%%%%%%%%%%%%%%%%%%%%%%%%%

%%%%%%%%%%%%%%%%% APPENDICES %%%%%%%%%%%%%%%%%%%%%

\appendix
\section{Results of the test predictions described in Section~5}

\begin{figure*}
	\includegraphics[width=1\textwidth]{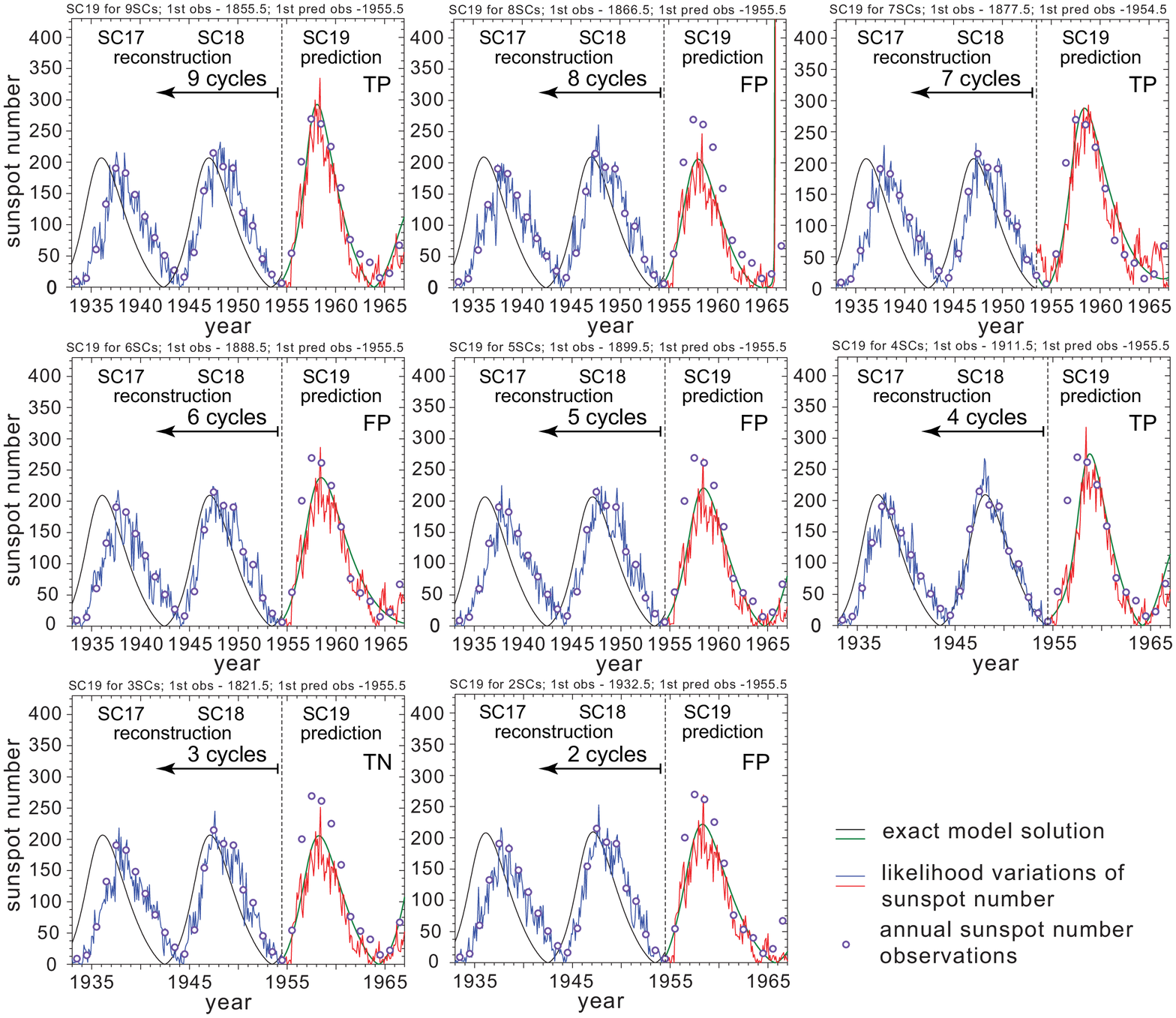}
	\caption{The test predictions of Solar Cycle 19 using the preceding sunspot number series of the various length, from 9 to 2 solar cycles. Black and green curves show the exact model solutions for the reconstruction and prediction phases. Blue and red curves correspond to the likelihood variations of the sunspot number. The actual annual sunspot number indicated by circles.}
	\label{fig:SC19}
\end{figure*}

\begin{figure*}
	\includegraphics[width=1\textwidth]{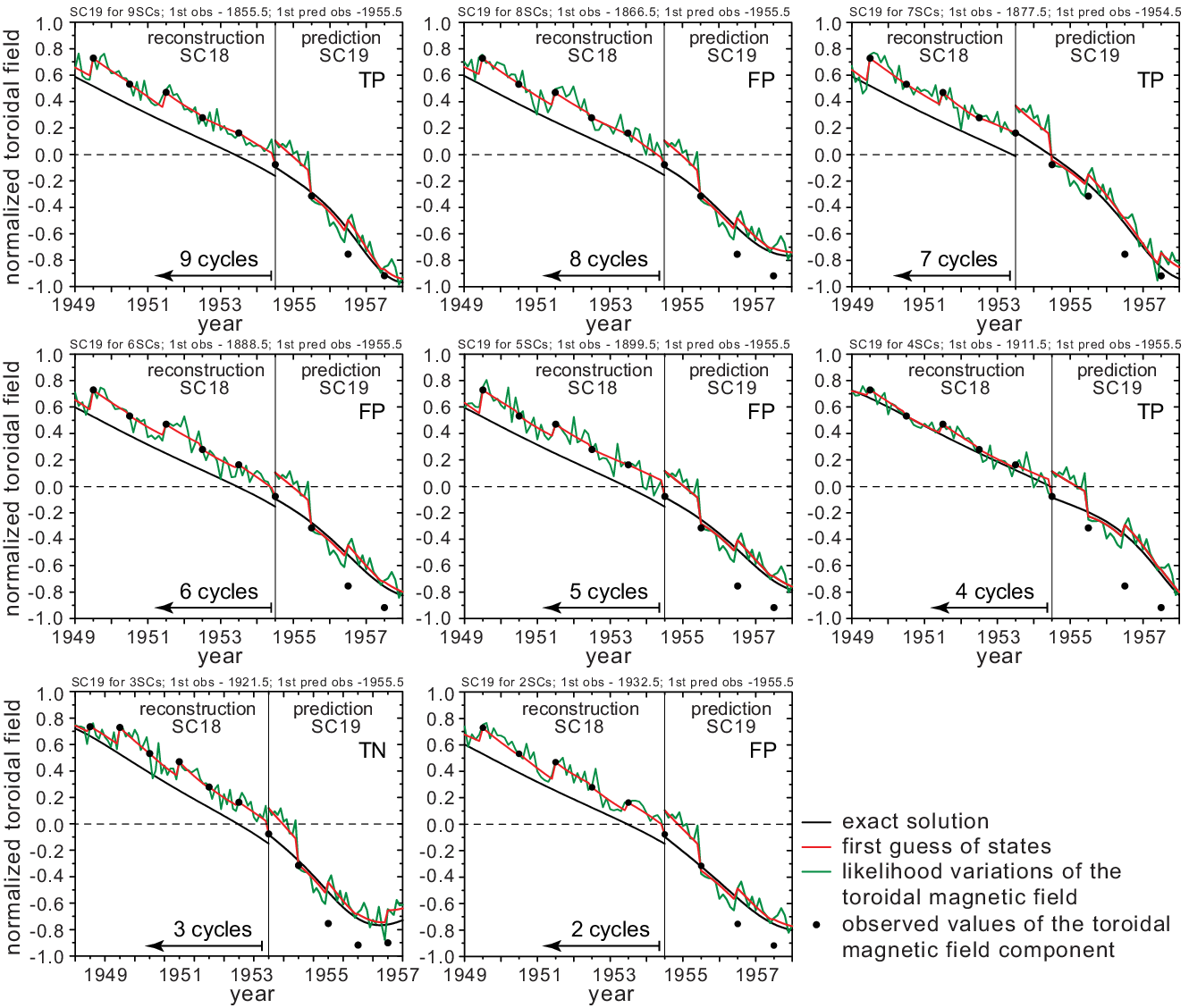}
	\caption{Solutions of the toroidal magnetic field for prediction of Solar Cycle 19 at the end of the reconstruction and the beginning of the prediction phases (1948-1957). Red curves indicate the corrected model solution (black curves), also called the first guess, according to observations the toroidal component of the magnetic field (dots).}
	\label{fig:SC19er}
\end{figure*}

\begin{figure*} 
	\includegraphics[width=1\textwidth]{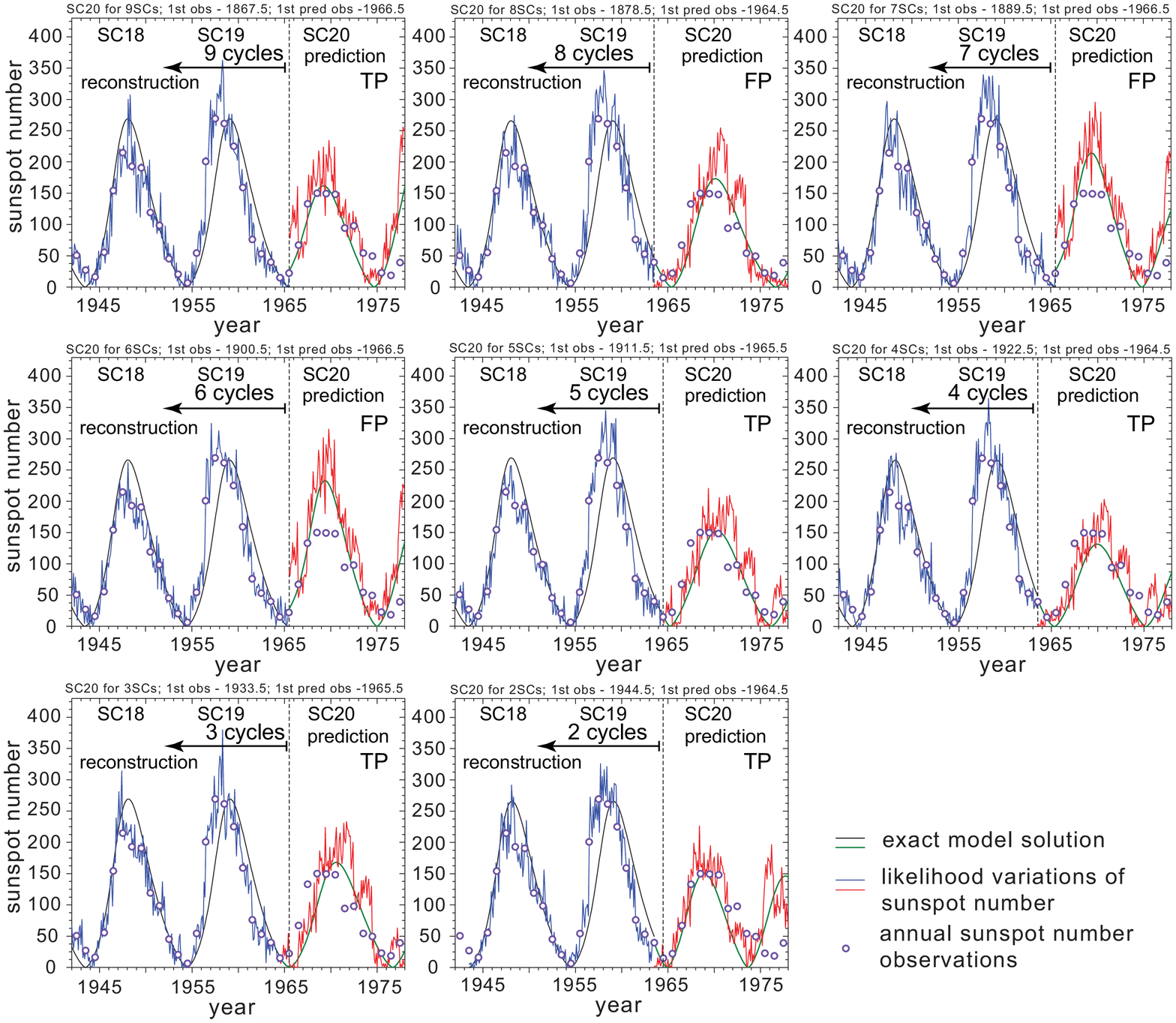}
	\caption{The same as in Figure~A1 %\ref{fig:SC19Pred9-2cycles} 
		for Solar Cycle 20.} \label{fig:SC20Pred9-2cycles}
\end{figure*}

\begin{figure*} 
	\includegraphics[width=1\textwidth]{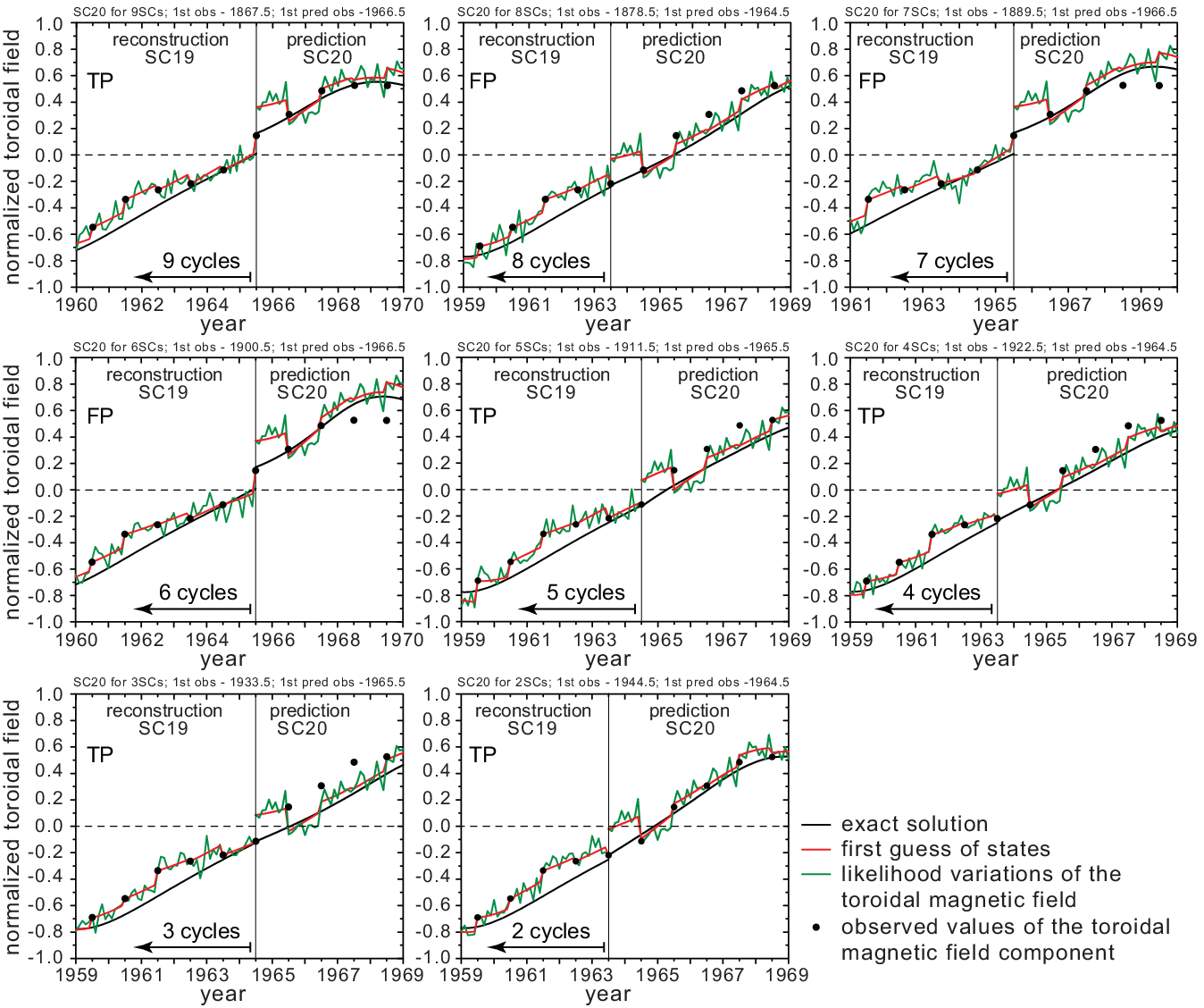}
	\caption{The same as in Figure~A2 %\ref{fig:SC19er}
		for Solar Cycle 20.} \label{fig:SC20er}
\end{figure*}

\begin{figure*} 
	\includegraphics[width=1\textwidth]{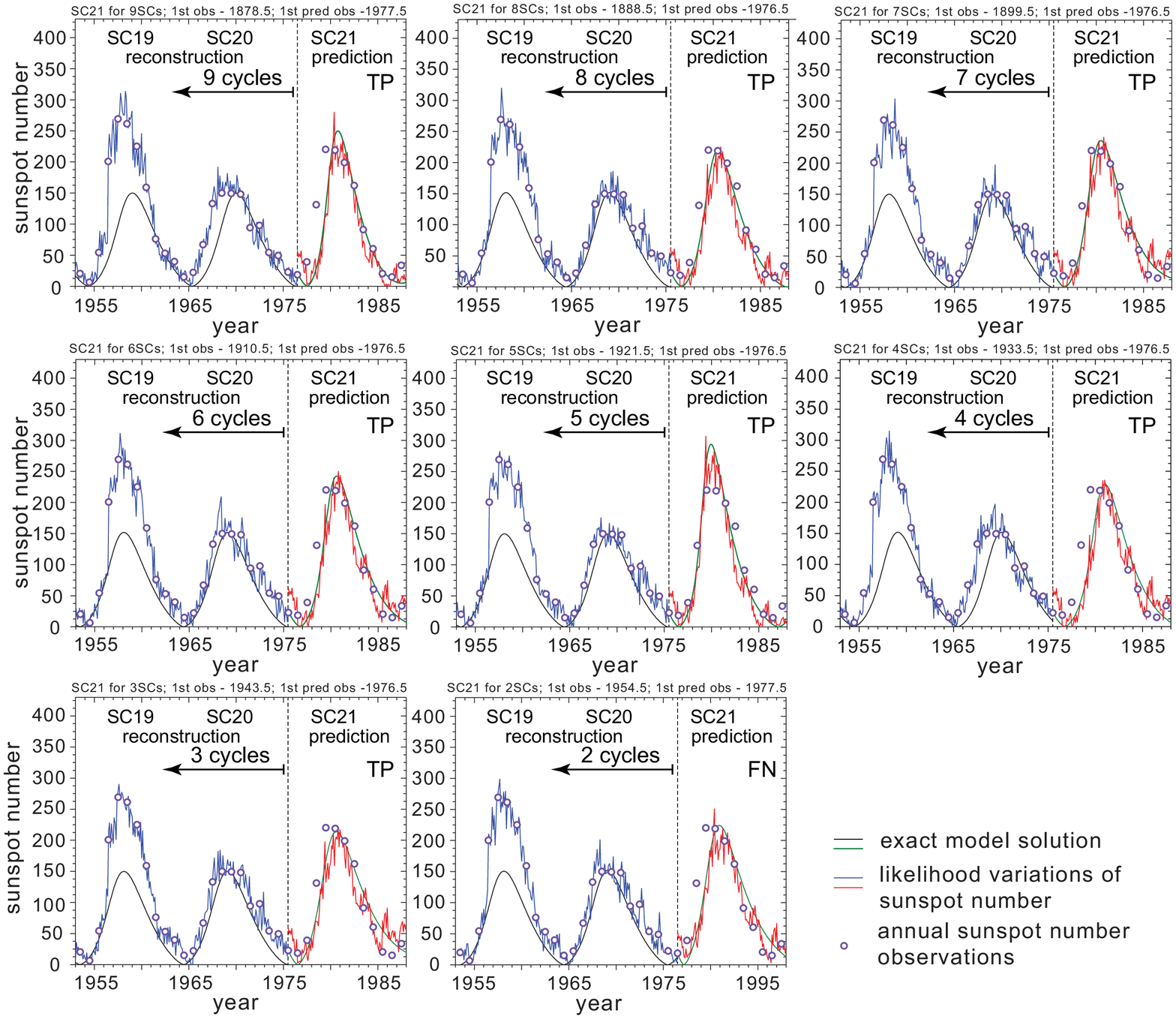}
	\caption{The same as in Figure~A1 %\ref{fig:SC19Pred9-2cycles}
		for Solar Cycle 21. }\label{fig:SC21Pred9-2cycles}
\end{figure*}

\begin{figure*} 
	\includegraphics[width=1\textwidth]{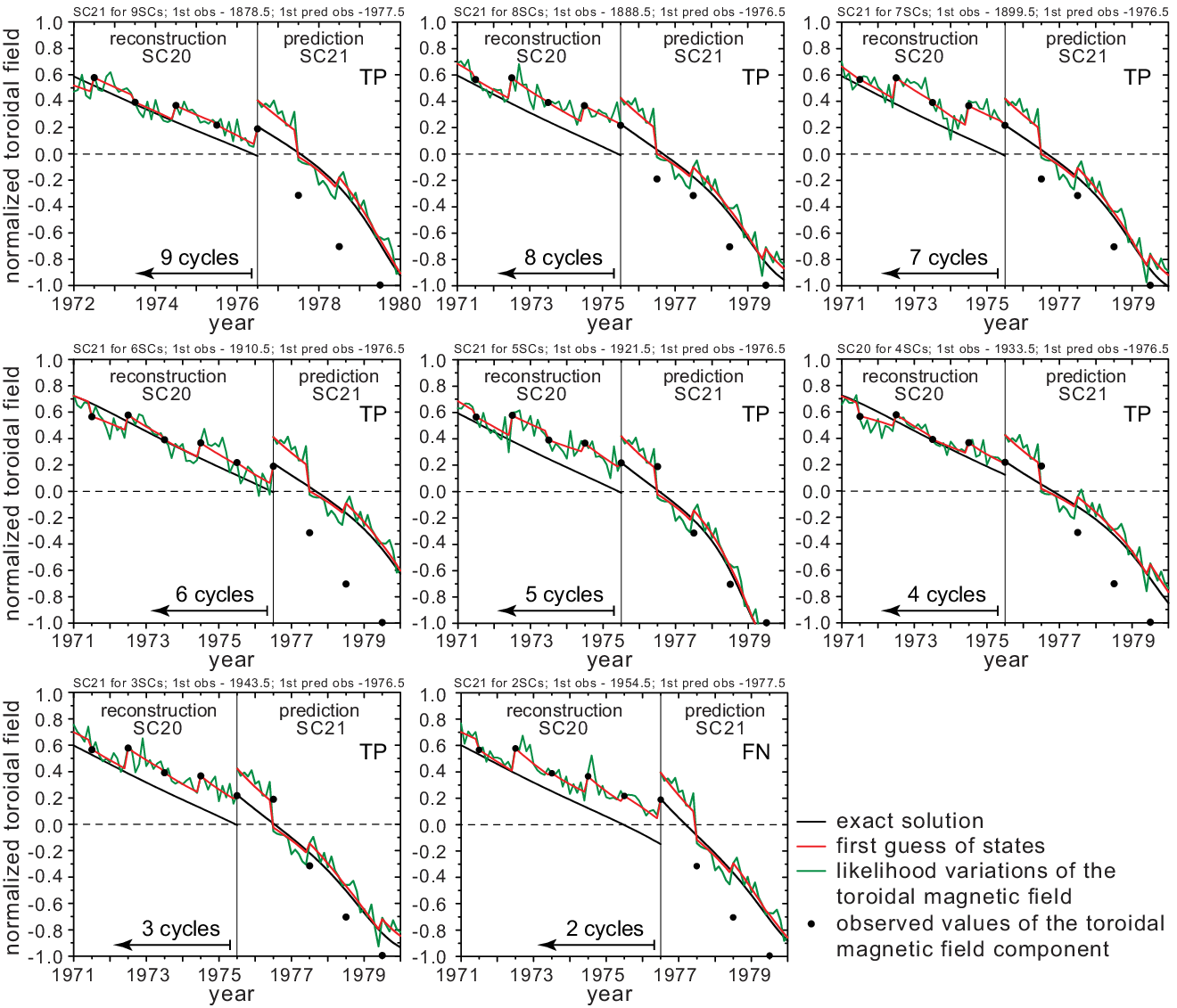}
	\caption{The same as in Figure~A2 %\ref{fig:SC19er}
		for Solar Cycle 21.} \label{fig:SC21er}
\end{figure*}

\begin{figure*} 
	\includegraphics[width=1\textwidth]{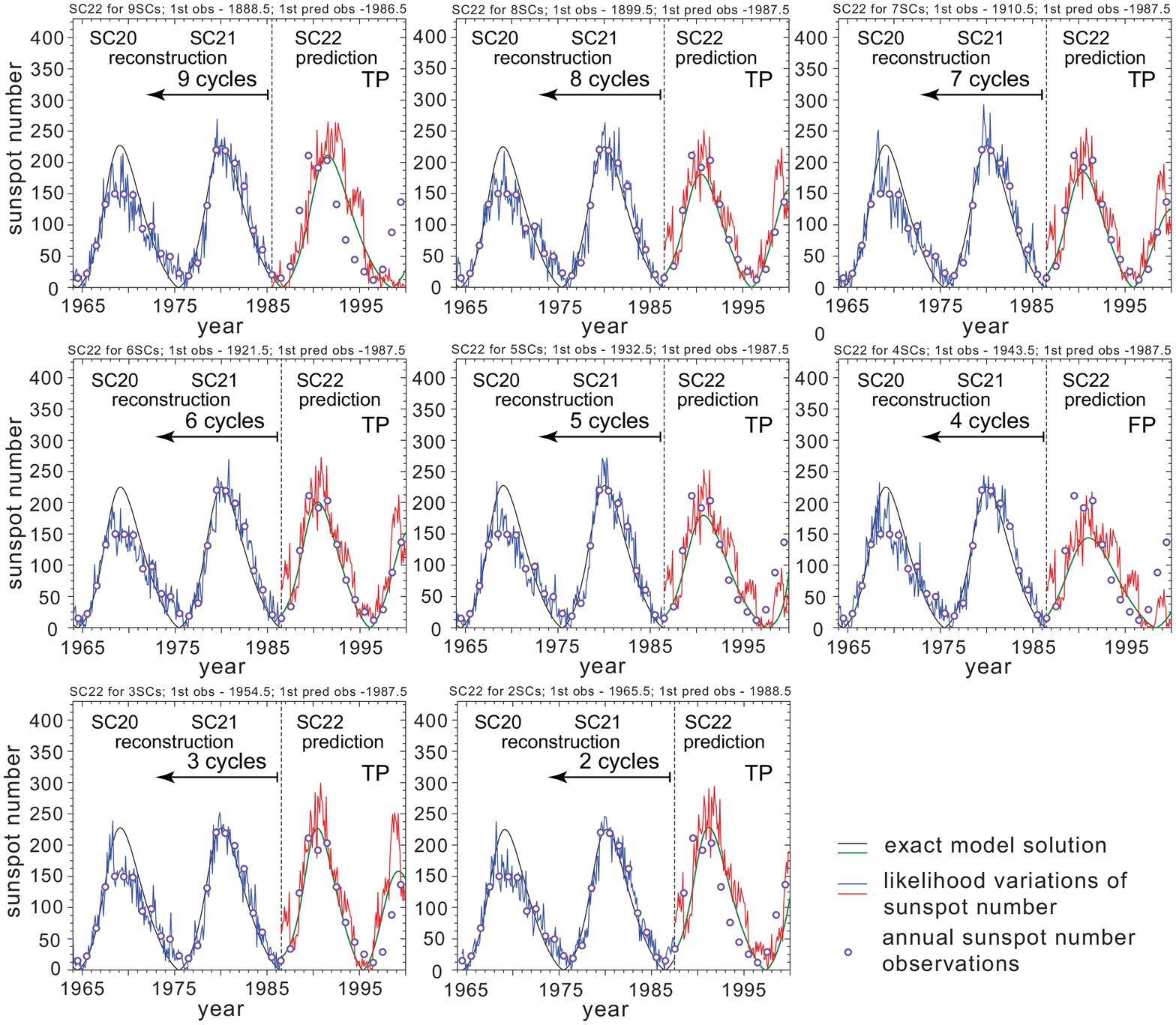}
	\caption{The same as in Figure~A1 %\ref{fig:SC19Pred9-2cycles}
		for Solar Cycle 22. }\label{fig:SC22Pred9-2cycles}
\end{figure*}

\begin{figure*} 
	\includegraphics[width=1\textwidth]{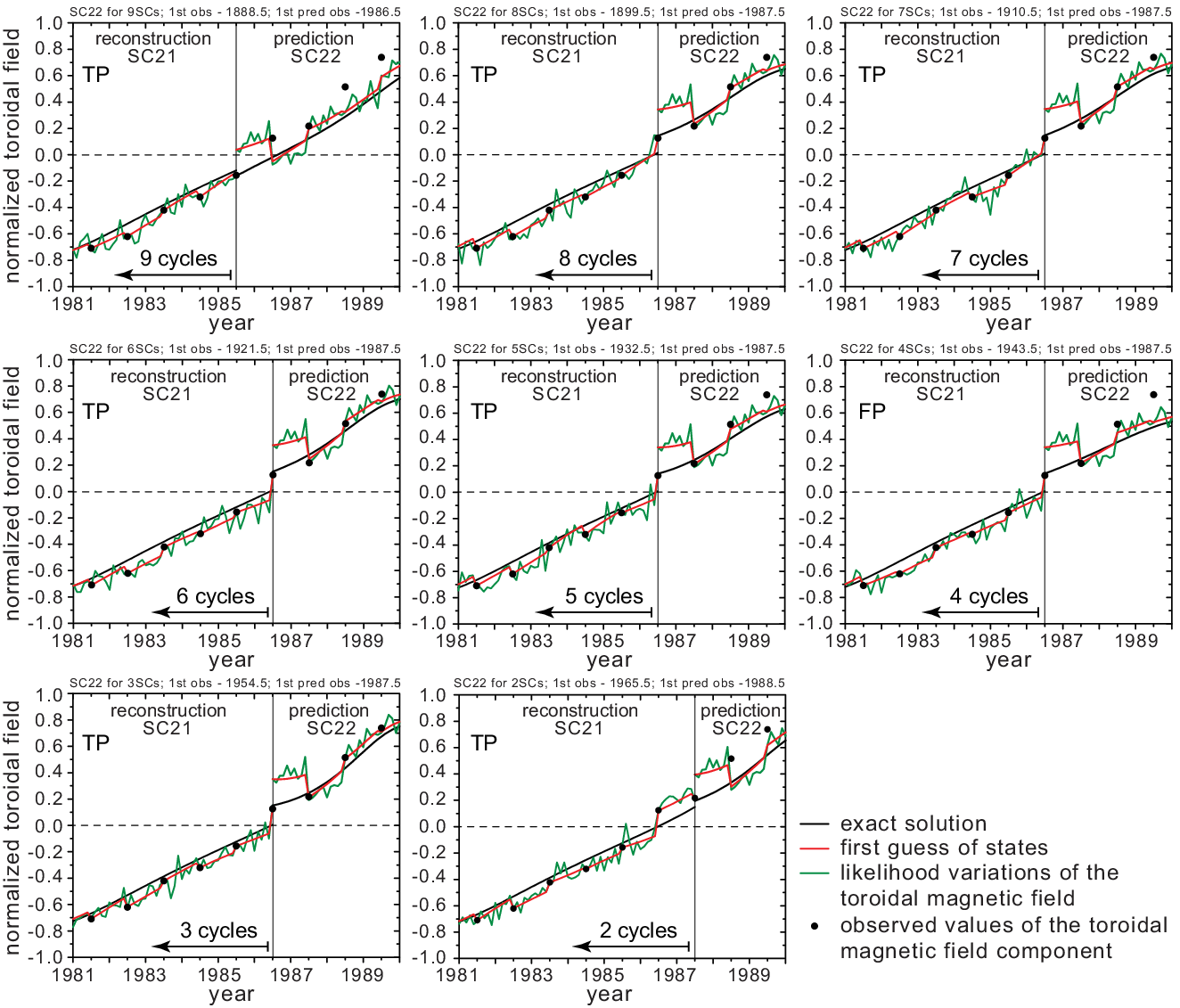}
	\caption{The same as in Figure~A2 %\ref{fig:SC19er} 
		for Solar Cycle 22.} \label{fig:SC22er}
\end{figure*}

\begin{figure*} 
	\includegraphics[width=1\textwidth]{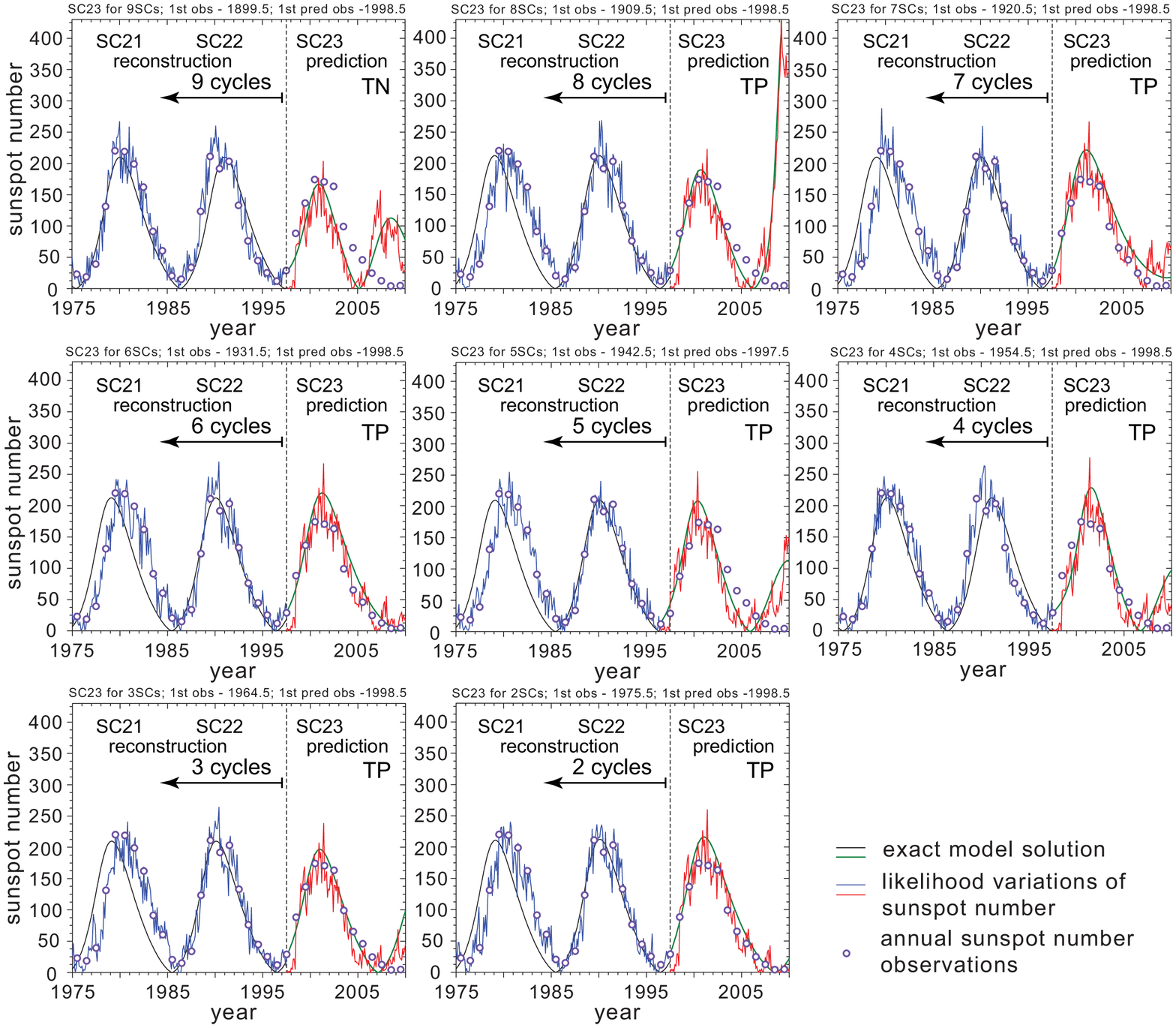}
	\caption{The same as in Figure~A1 %\ref{fig:SC19Pred9-2cycles}
		for Solar Cycle 23.} \label{fig:SC23Pred9-2cycles}
\end{figure*}

\begin{figure*} 
	\includegraphics[width=1\textwidth]{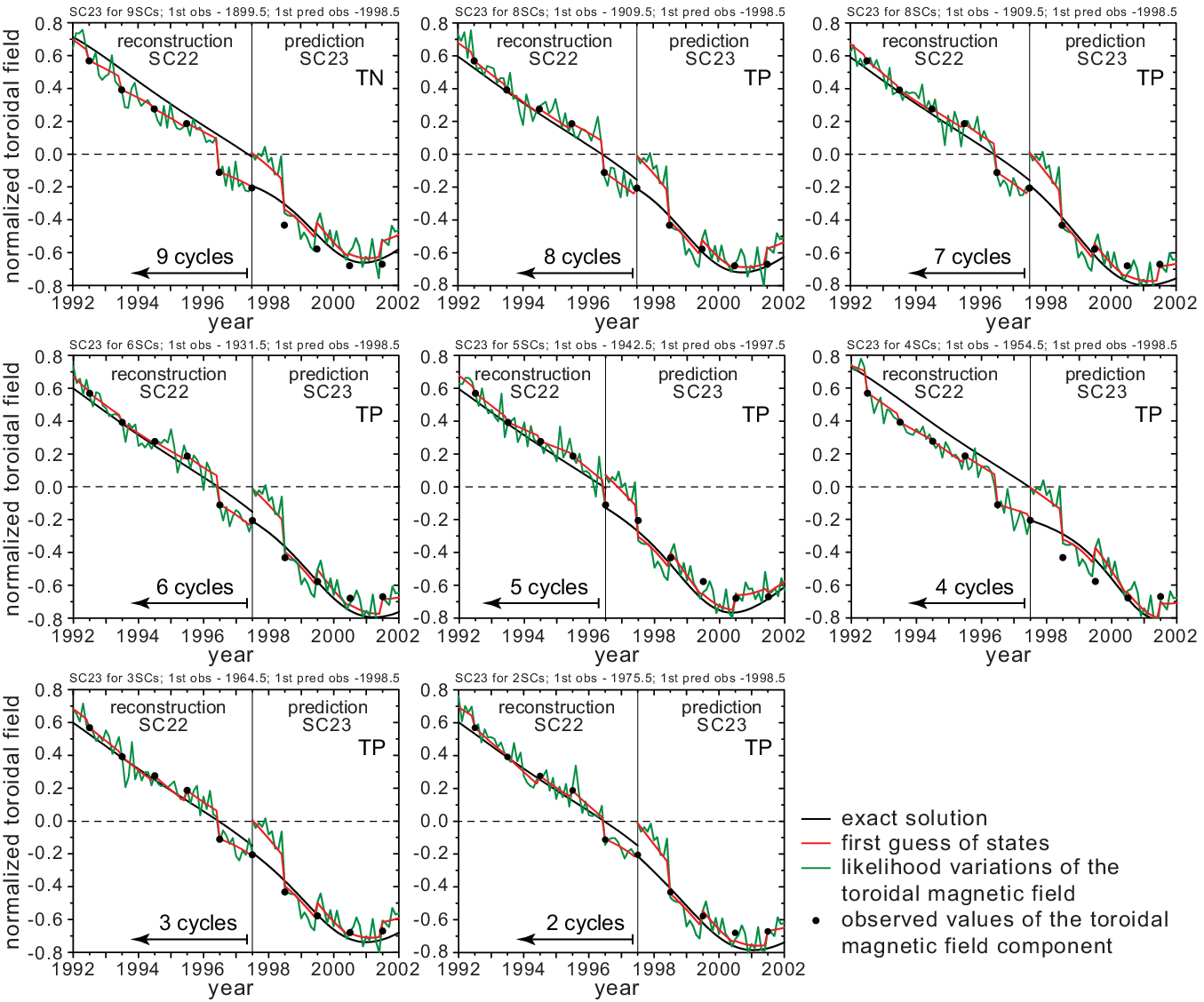}
	\caption{The same as in Figure~A2 % \ref{fig:SC19er}
		for Solar Cycle 23.} \label{fig:SC23er}
\end{figure*}

\begin{figure*} 
	\includegraphics[width=1\textwidth]{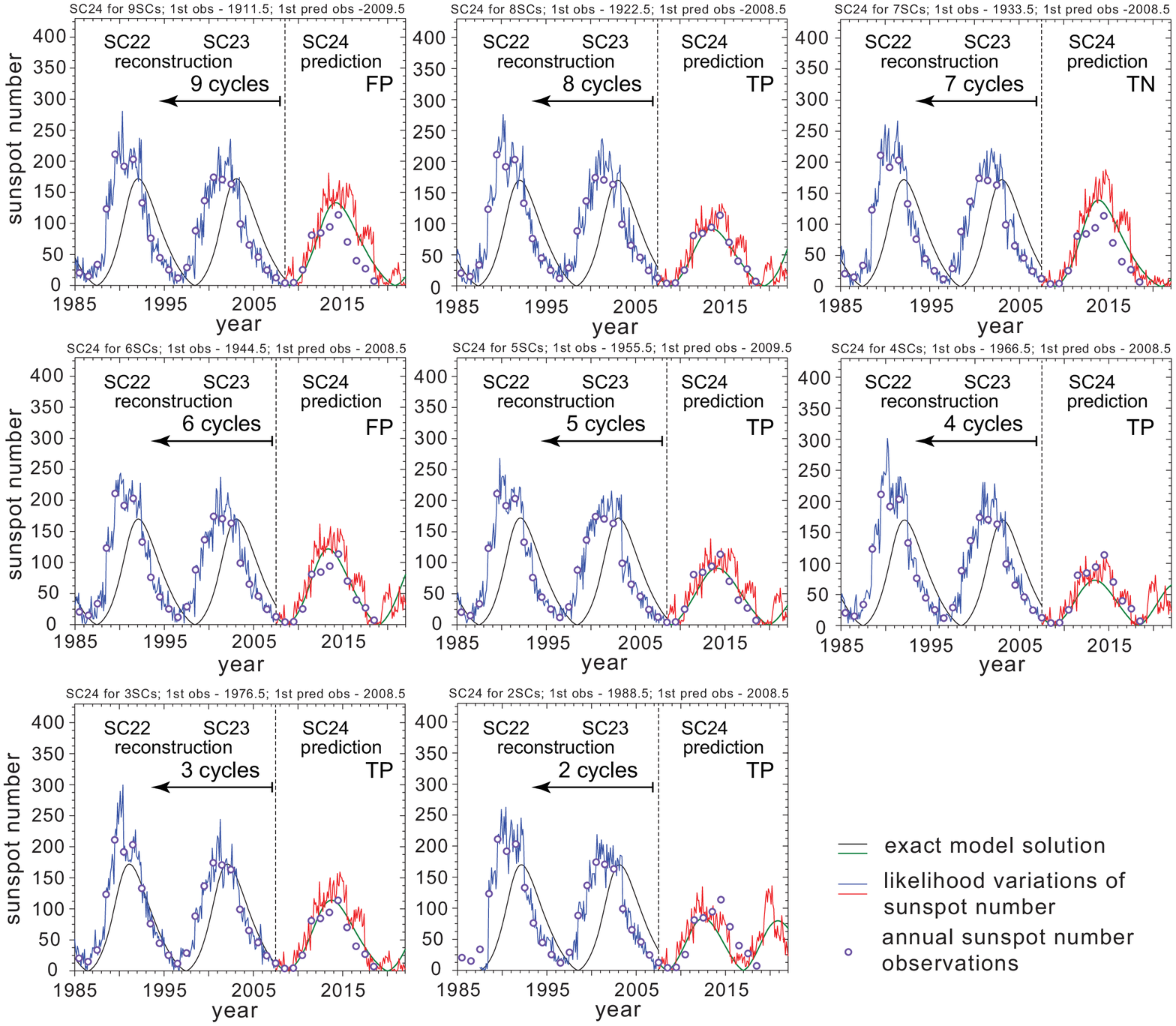}
	\caption{The same as in Figure~A1 %\ref{fig:SC19Pred9-2cycles}
		for Solar Cycle 24. }\label{fig:SC24Pred9-2cycles}
\end{figure*}

\begin{figure*} 
	\includegraphics[width=1\textwidth]{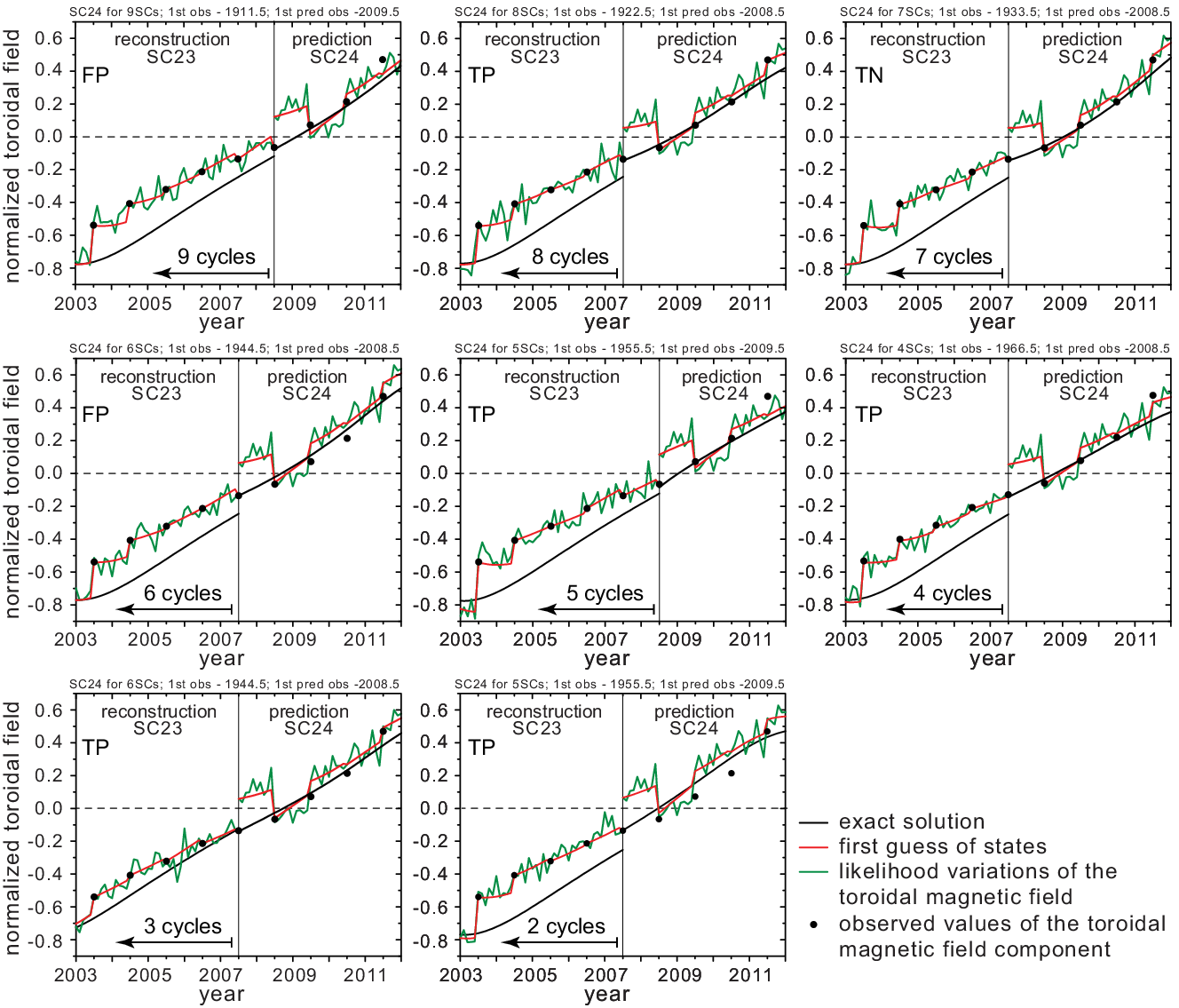}
	\caption{The same as in Figure~A2 % \ref{fig:SC19er}
		for Solar Cycle 24.} \label{fig:SC24er}
\end{figure*}

%%%%%%%%%%%%%%%%%%%%%%%%%%%%%%%%%%%%%%%%%%%%%%%%%%

% Don't change these lines
\bsp	% typesetting comment
\label{lastpage}
\end{document}